\documentclass[conference]{IEEEtran}
\IEEEoverridecommandlockouts
\usepackage{cite}
\usepackage{amsmath,amssymb,amsfonts}
\usepackage{algorithmic}
\usepackage{graphicx}
\usepackage{textcomp}
\usepackage{xcolor}
\usepackage{multicol}
\usepackage{epstopdf}
\def\BibTeX{{\rm B\kern-.05em{\sc i\kern-.025em b}\kern-.08em
    T\kern-.1667em\lower.7ex\hbox{E}\kern-.125emX}}
\begin{document}

\title{Optimizing Furnace efficiency for Factory of Future using Cooperative Games\\
}

\author{\IEEEauthorblockN{Sreenath Shaju, Mohak Sukhwani, Ankit Kala}
\IEEEauthorblockA{ABB Industrial Automation and Digitalization, India\\
sreenath.s@in.abb.com, mohak.sukhwani@in.abb.com, ankit.kala@in.abb.com\\
}}

\maketitle

\begin{abstract}
Approximately 65 to 90 percent of energy used in petrochemical and refining industries is consumed by furnaces. Operating furnaces at optimal conditions results in huge amounts of savings in energy consumption. In this paper, we model the furnace efficiency optimization as a multi-objective problem involving multiple interactions among the controlled variables and propose a cooperative game based formulation for the factory of future. The controlled variables are Absorbed Duty and Coil Outlet Temperature. We propose a comprehensive solution to select the best combination of manipulated variables (fired duty, throughput and coil inlet temperature) satisfying multiple criteria using a cooperative game theory approach. We compare this approach with the standard multi-objective optimization using ~\textsc{NSGA-II} and ~\textsc{RNSGA-II} algorithms. The new approach provides a single optimum solution, instead of a set of equally efficient configurations, as in the case of ~\textsc{NSGA-II} and ~\textsc{RNSGA-II} algorithms.
\end{abstract}

\section{Introduction}
The trends in manufacturing have gone through major shifts in last few years. Disruption in information and operational technologies along with substantial advancements in AI have led to this paradigm shift. Traditional factory floors are looking for sustainable AI solutions to increase their productions with considerable reduction in waste and carbon footprint. The Factory of the Future is essentially about solving this demand, automation using real time factory data for high quality throughput, optimized control and higher utilization rates. It is a collection of advanced digital technologies like internet of things, cognitive computing and cyber-physical systems working in tandem to produce desired results. We propose one such, game theory driven, plant process optimization method for efficient furnace operation. 

There are multiple heating and cooling processes associated with chemical industries and in majority of these processes, it might be necessary to heat the process fluids to a certain temperature. This is generally achieved using furnaces. Furnace is typically a kind of heat exchanger that transfers the thermal energy obtained from combustion of fuels in a closed space to the process fluid. It is a crucial component of any refinery or petrochemical process unit, primarily used to crack or heat process fluids for separation/conversion of crude, heavy oils and low value products to higher valuable products. Typically furnaces are part of process units, viz. distillation, treatment, conversion etc. A furnace on an average consumes ~65-90 $\%$ \cite{furnace} of the total energy and hence optimizing furnace performance is of utmost importance. 

We consider an induced draft furnace which operates by virtue of high temperature difference between the burner and the stack. There are numerous opportunities to optimize its performance. We first investigate the scope for optimization. In a furnace, the temperature of the process fluid at the inlet (Coil Inlet Temperature, typically called CIT) is increased to a temperature (Coil outlet Temperature, typically called COT), till it cracks and gets converted into valuable products and by-products as a result of the heat energy released from combustion of fuels (fired duty). The value of COT is typically targeted based on the downstream process (such as a column or a reactor) requirements or furnace mechanical design limits. As in any furnace, the total fired duty is not completely utilised by the process fluid, only a fraction of it is used and that fraction is called absorbed duty. Both the fired and absorbed duties are directly proportional to the temperature difference across the furnace. To achieve a higher temperature difference, it is important to maximize the values of COT and absorbed duty.

\begin{figure}[t]
    \centering
    \includegraphics[scale=0.34]{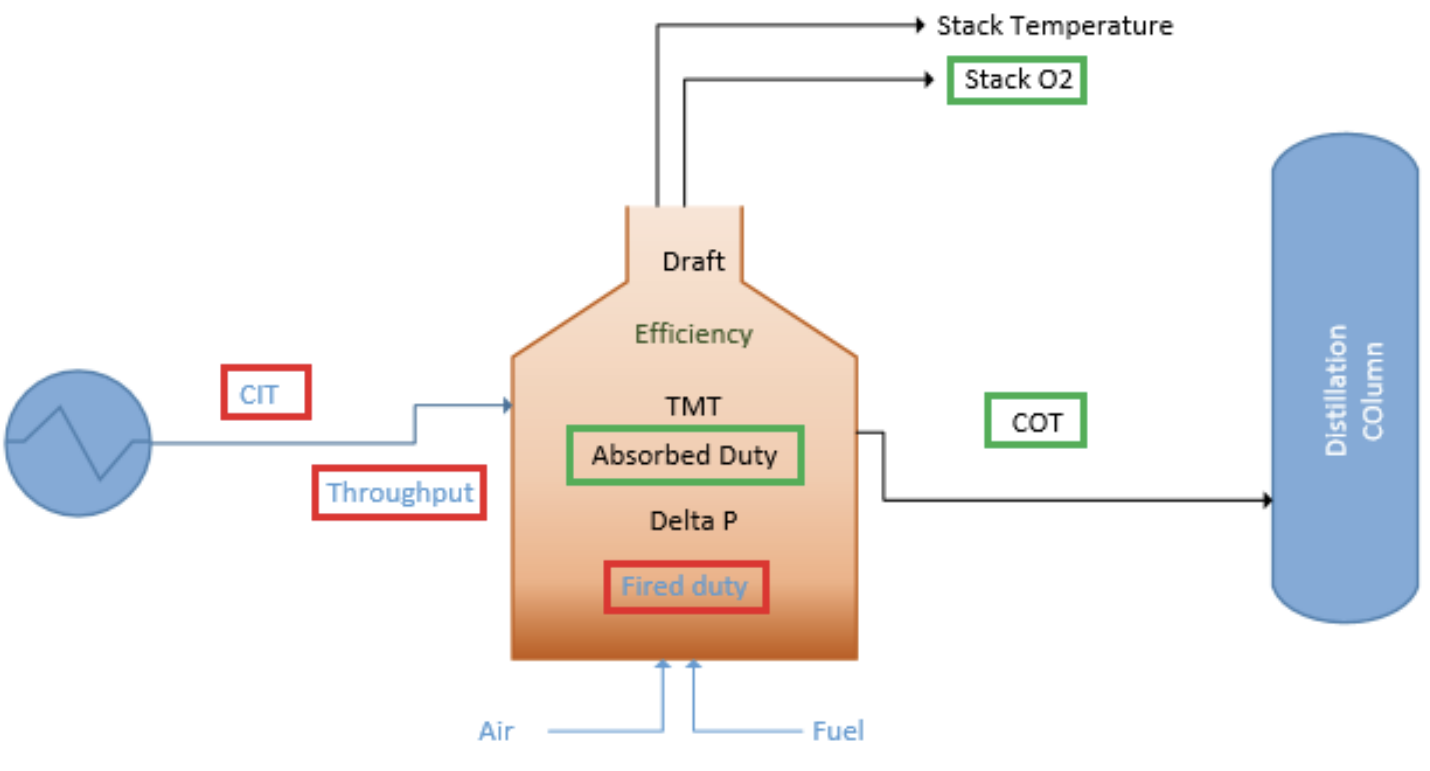}
    \caption{Illustration of a typical furnace in a process unit: CIT, Throughput, Fired Duty are the identified manipulated variables and Stack O2, COT and Absorbed duty are the controlled variables. Statistical modeling based approaches are used to study the effect manipulated variables on the controlled variables. The model outputs are used to perform optimization of furnace by formulating it as a cooperative game based interaction among trained models.}
    \label{fig:intro}
\end{figure}

The burning process which requires air should also be subjected to optimization. The excess air results in oxygen that isn’t consumed during combustion, and this oxygen absorbs otherwise usable heat and carries it out of the stack. The chemically ideal amount of air entering a furnace is just enough for all the oxygen in the air to be consumed. However, this ideal proportion (known as the stoichiometric air-to-fuel ratio) is difficult to reach because fuel and air do not completely mix, which means, a certain amount of excess air will always be necessary for complete combustion. In fact, too little excess air results in inefficient burning of fuel, CO production, and unnecessary greenhouse gas emissions resulting in undesirable effects on the environment. 

Hence, it is clear that ensuring optimization of furnace efficiency is a complex task because it involves interactions among multiple monitoring parameters such as Coil Inlet Temperature (CIT), Process flow (Throughput), air and fuel parameters, draft, Tube Metal Temperature (TMT), Coil Outlet Temperature (COT), Stack temperature and oxygen etc. Many calculated parameters such as Fired duty, Absorbed duty, Efficiency are used as metrics to optimize its performance, (see figure~\ref{fig:intro}). In a traditional setup, a process engineer (or an operator) typically optimizes a furnace performance by changing these variables based on his operating experience, thumb rules, guidelines, design and process calculations. However in context of Factory of future, artificial intelligence techniques such as machine learning, game theory and optimization can be employed to decide values for these set-points. 

\begin{table*}[h!]
  \centering
	\begin{tabular}{cccccccc}
	
	\hline
	
	Manipulated Variables &  Controlled Variables & Train MSE &  Test MSE &  Train RMSE &  Test RMSE &  Train Rsquare  &  Test Rsquare \\	
	\hline
	Fired Duty, Throughput, CIT
       &  Absorbed Duty
 &             3.523 &        5.105 &                        1.877 &         2.259 &       0.941 &                     0.925 \\
	Fired Duty, Throughput, CIT       &  Stack O2 &             0.106 &        0.142 &                        0.326 &         0.377 &       0.349 &                     0.207 \\
	Fired Duty, Throughput, CIT       &  COT &             11.505 &        21.135 &                        3.392 &         4.597 &       0.810 &                     0.687  \\
	\hline
\end{tabular}
\vspace{0.3cm}
\caption{Decision Tree Models - Performance Measures}
\label{dt}
\end{table*}

With the cooperative game theory driven approach, we assume that the furnace efficiency is a result of multiple agents working in tandem to achieve their own individual goals. These agents are conveniently called controlled variables, because these are the variables that indicate the optimality of the process. The controlled variables are functions of another set of variables known as manipulated variables. Preliminary analysis of furnace as a system indicates that CIT, Fired Duty and Throughput are the manipulated variables and Absorbed Duty, stack oxygen and COT are the controlled variables. Further the controlled variables may get filtered depending on fitness of the models built for these variables. Figure~\ref{fig:intro} illustrates the identified manipulated variables and controlled variables. Each controlled variable (in green box) is a function of the manipulated variables. This function is captured using machine learning models which in turn participate in a cooperative game. 

We are ideally in search of the best configuration of manipulated variables that lead to furnace efficiency optimization. This often results in multi-objective optimization problems, such as maximizing the absorbed duty, minimizing the excess oxygen, maximizing the coil outlet temperature etc. It is difficult to obtain a single solution to the multi-objective optimization problem because improvement in one objective comes at the expense of another. These problems are usually solved using evolutionary algorithms such as ~\textsc{NSGA-II}, ~\textsc{NSGA-III}, ~\textsc{EFRRR} etc \cite{osti_1529693}. These techniques result in a set of optimal solutions which is usually called Pareto Optimal set. All the solutions in this set are good, but decision making is still a question because we do not get a single configuration of the best manipulated variables. Instead, we get a set of equally efficient configurations. The inherent limitation which still creates confusion on what value to choose should be overcome by an expert or (for a truly automated scenarios) there should be an automated way to capture multiple interactions among the controlled variables. We address this problem of automation using cooperative games. 

\section{Related Work}
Furnace optimization has been a problem of great interest and several approaches such as mixed integer programming have been used to achieve this. With respect to optimal design of furnace, a detailed Mixed Integer Nonlinear Programming (MINLP) model including operational and geometric constraints was developed by the authors in \cite{MUSSATI20092194}.

Game Theory has been used extensively to solve  optimization problems, not particularly to furnace optimization scenario. With respect to multi-objective optimization, several approaches based on both cooperative as well as non-cooperative game theory have been proposed and studied. In \cite{doi:10.1080/03052150902890064}, a game-theory approach has been used for the multi-objective optimum design of stationary flat-plate solar collectors. Game theoretical idea is used to solve tri-objective constrained optimization problems to find a balanced solution. In \cite{RAO1987119}, the authors develop a computational procedure to solve a general multi-objective optimization problem using cooperative game theory and multiple numerical examples are used to illustrate this. The authors in \cite{8680809} propose a cooperative game based mathematical model of an intelligent multi-objective home energy management (HEM) scheme with the integration of small-scale renewable energy sources.  In this game, each HEM objective is assigned as a player and every player tries to maximise their own payoff.

In \cite{article}, the authors provide a brief and self-contained introduction to the theory of cooperative games and explains how this can be used to get acquainted with the basics of cooperative games. The authors analyze cooperative game theory in the
context of environmental problems and models. The main models (bargaining games, transfer utility and non transfer utility games) and issues and solutions are considered: bargaining solutions, single-value solutions like the Shapley value and the nucleolus, and
multi-value solutions such as the core are explained in depth. 

The paper \cite{5554307} presents the game description of multi-objective optimization design problem and takes the design objectives as different players. The design variables are divided into different strategic spaces owned by each player by calculating the affecting factors of the design variables to objective functions and fuzzy clustering. It uses Nash equilibrium game model, coalition cooperative game model and evolutionary game model to solve multi-objective optimization design problem and gives corresponding solving steps. 

In \cite{pub.1101630721}, a multi-objective leader-follower game based on the Stackelberg model, where the designer’s preferred target is taken into account and a real-life example of the multi-objective optimization design of a Chinese arch dam named “Baihetan” is presented to demonstrate the effectiveness of the proposed method.
The authors in \cite{pub.1042336139} propose a multi-objective optimization method based on self-adaptive space division of design variables and carry out tri-objective optimization for vehicle suspension using this method and it is compared with the traditional game method. 

Application of game theory for solving optimization problems, has been thus well studied. We extend the application of game theory towards process optimization. The main contributions of this paper are summarised as follows:
\begin{itemize}
    \item Solve the multi-objective optimization problem for furnace efficiency optimization using \textsc{NSGA-II} and \textsc{RNSGA-II} algorithms, see section \ref{nsga2})
    \item Provide a comprehensive methodology to achieve optimal values of manipulated variables using cooperative game theory via Nash Bargaining and collaborative negotiations (see section \ref{copgame}) and implement the same
    \item Compare the solutions of the Nash Bargaining Model with the \textsc{NSGA-II} and \textsc{RNSGA-II} results
\end{itemize}

\section{Problem Description}
We consider a draft type furnace data, which consists of multiple feature variables such as Timestamp, Stack-O2 (stack oxygen or excess oxygen in \%), Efficiency (furnace efficiency in \%), Fuel-Gas (fuel gas flow rate in kg/hr), Fired-duty-MW (fired duty in MW), Absorbed-duty-MW (absorbed duty in MW), Throughput, CIT-degC (coil inlet temperature in celsius) and COT-degC (coil outlet temperature), for our experimentation. The variables whose values can be manipulated are Fired Duty, Throughput and CIT. The values of these variables can be manipulated in order to control the values of Absorbed Duty, Stack O2 and COT. To improve furnace efficiency, we primarily focus on maximizing Absorbed Duty and COT and maintaining the Stack O2 value between between 1.5 $\%$ and 2 $\%$.

\begin{figure}
    \centering
    \includegraphics[scale=0.37]{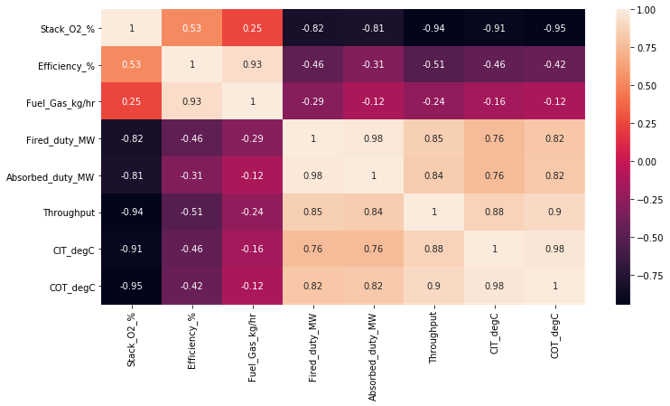}
    \caption{Correlation Matrix capturing relations among all the variables in the data set. Both Absorbed duty and COT are strongly correlated with Fired Duty, Throughput and CIT.
}
    \label{cor}
\end{figure}

From the systemic view of furnace, it is clear that the controlled variables are independent of each other, but each of them depends on the three manipulated variables. Figure ~\ref{cor}, shows the correlation values of the variables in the data set. It is evident that three models would be required to predict the three controlled variables as functions of the three manipulated variables. Accordingly, regression models were developed for each of the three controlled variables using decision trees. Table~\ref{dt}, shows the performance measures of these models. Based on the results of the regression models, we discard the stack oxygen model as its performance is worst. We consider the other two models and formulate multi objective optimization problem using Absorbed Duty model and COT model.

\subsection{Multi-Objective Optimization}\label{nsga2}
The furnace optimization problem has been reduced to a bi-objective optimization problem which aims to maximizing both Absorbed Duty and COT, subject to bounds on the manipulated variables. The bounds on the manipulated variables are decided based on the operating conditions as well as values in the data set. Let $x_1$, $x_2$ and $x_3$ be the manipulated variables, Fired Duty, Throughput and CIT respectively. Let $Y_1$, $Y_2$ be the controlled variables,  Absorbed Duty and COT respectively. Objectives of the optimization model is to maximize the Absorbed Duty and COT as shown in equations (1) and (2). 

\begin{figure}[b]
    \centering
    \includegraphics[scale=1.1]{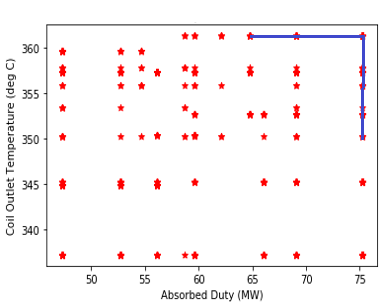}
    \caption{Feasible Solutions and Pareto-Optimal Frontier. The blue line shows the pareto-optimal frontier and the feasible solutions that lie on this frontier maximize both Absorbed Duty (MW) as well as Coil Outlet Temperature (deg C).}
    \label{pareto}
\end{figure}

\begin{align} 
   J_1=\text{max } Y_1 \\
   J_2=\text{max } Y_2 
\end{align}

Constraints (3), (4) and (5) are the bounds on the manipulated variables. The lower bound on CIT ($x_3$) was defined as the lowest achieved temperature (based on the operating data) whereas the upper bound was decided based on the mechanical design temperature of downstream equipment. It is assumed that the downstream process is not constrained by temperature (constraints such as coking, cracking etc. in downstream equipment). With respect to throughput ($x_2$), the lower bound was decided based on unit turn-down limit, which is defined as the lowest throughput at which a unit can operate with final products within quality limits. The upper bound was decided based on unit (maximum) design throughput. The lower bound on fired duty ($x_1$) was defined based on the furnace (burner) turn-down limit and upper bound was decided based on furnace (maximum) design firing limit. 
\begin{align} 
   44.4 \leq x_1 \leq 103 \\
   58.6 \leq x_2 \leq 107 \\
   176.3 \leq x_3 \leq 223
\end{align}

The optimization model was solved using two genetic algorithms, viz. ~\textsc{NSGA-$II$} \cite{996017} and ~\textsc{RNSGA-$II$} \cite{10.1145/1389095.1389231}, using pymoo~\cite{pymoo} in Python. For~ \textsc{NSGA-$II$}, the following parameters were used:

\begin{multicols}{2}
\begin{itemize}
    \item Population size: 40
    \item Offspring size: 10
    \item Crossover Type: SBX~\footnote{Simulated Binary Crossover}
    \item Crossover Prob : 0.9
    \item $\eta_{crossover}$ : 15
    \item Mutation: Polynomial
    \item Mutation Prob : 0.1
    \item $\eta_{mutation}$ : 20
\end{itemize}
\end{multicols}

The following parameters were used to solve the problem using \textsc{RNSGA-$II$}.

\begin{itemize}
    \item Population size: 40
    \item Reference Points: (40, 90) and (10, 278)
    \item $\epsilon$: 0.01
    \item Weights of objectives: (0.5, 0.5)
\end{itemize}

Pareto-Optimal configurations were obtained as part of the optimization. Figure (\ref{pareto}) shows the feasible solutions and the pareto-optimal frontier (shown in blue line). There are multiple configurations of manipulated variables that give the optimal values for the objectives. Let $Y_1^*$ and $Y_2^*$ be the optimal values of Absorbed Duty and COT respectively. These values are same for both the algorithms. Based on \textsc{NSGA-II} and \textsc{RNSGA-II} results, these values are:
\begin{equation} \label{nsga}
    (Y_1^*, Y_2^*) = (75.20, 361.29)
\end{equation} 
Thus, the maximum values of absorbed duty in MW and COT in $^oC$ are $75.20$ and $361.29$ respectively.

\subsection{Cooperative Games for Optimizing Furnace Efficiency} \label{copgame}
With respect to furnace optimization, two objective functions are assumed as two different players, one for Absorbed Duty and other for COT. This is then modelled as a two person Nash Bargaining Problem, ~\cite{8680809}. First individual optimization problems are solved for both the objectives to identify the best and worst solutions. The worst solution is the disagreement point for any player, because this is the least payoff any player can get once they enter the game. With the worst solution as reference, the players bargain to get better solutions. 

The optimization problem is formulated as:
\begin{equation}
   \begin{aligned} \label{ab_Opt}
   J_1 = \text{max } Y_1 \\
   \text{subject to constraints:} \\
   44.4 \leq x_1 \leq 103 \\
   58.6 \leq x_2 \leq 107 \\
   176.3 \leq x_3 \leq 223
\end{aligned} 
\end{equation}

The optimization problem was solved to get the best solution $X_{1_{best}}$ that maximises $Y_1$. Let, $Y_{1_{best}}$ be the best value of the absorbed duty obtained. It is observed that,

\begin{align}
 X_{1_{best}} &= (86.18, 101.71, 176.88),\\
 Y_{1_{best}} &= 75.2
\end{align}

We solve the individual optimization problem to maximize COT via Genetic Algorithm using pymoo~\cite{pymoo}. The optimization problem is formulated as:

\begin{equation}
\begin{aligned} 
   J_2 = \text{max } Y_2,    \\
\text{subject to constraints:} \\
   44.4 \leq x_1 \leq 103 \\
   58.6 \leq x_2 \leq 107 \\
   176.3 \leq x_3 \leq 223
\end{aligned}
\end{equation}

Following the similar approach, we get the best values for COT by solving this optimization. Let, $Y_{2_{best}}$ be the best value of COT obtained and $X_{2_{best}}$ be the solution of manipulated variables. It is observed that,

\begin{align}
    X_{2_{best}} &= (77.27, 95.22, 213.69) \\
    Y_{2_{best}} &= 361.29
\end{align}


\begin{figure*}[h!]
    \centering
    \includegraphics[scale=0.64]{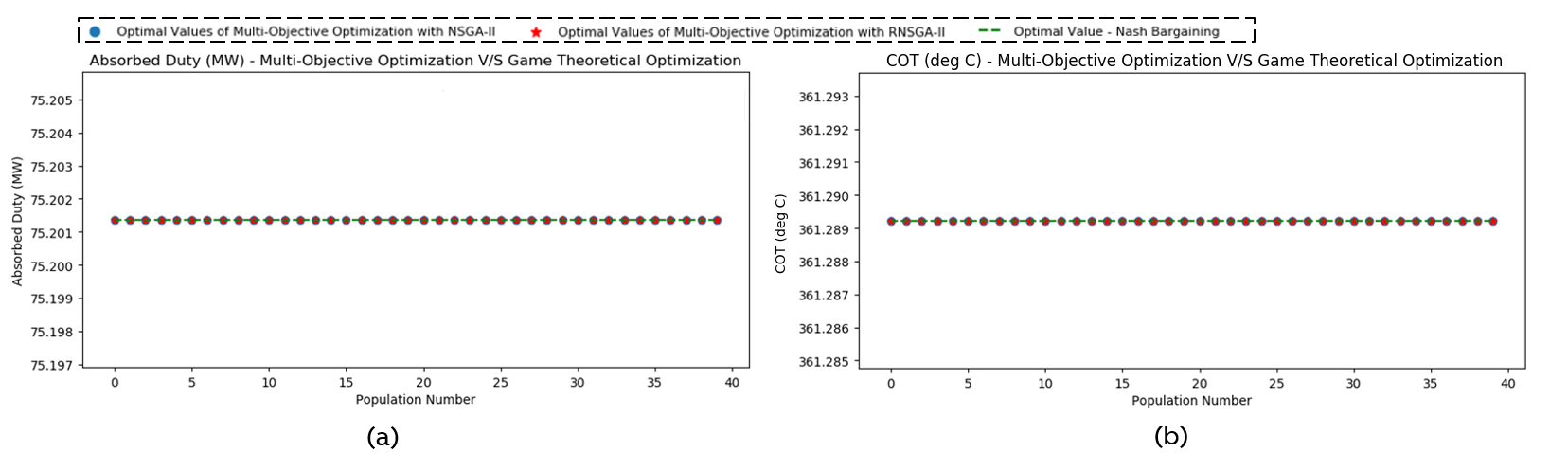}
    \caption{Comparison of optimal values of the controlled variables (objective functions) obtained via Nash Bargaining, \textsc{NSGA-II} and \textsc{RNSGA-II} algorithms. The green dotted line shows the Nash Bargaining solution whereas the blue dots and the red stars indicate the optimal values of each population via \textsc{NSGA-II} and \textsc{RNSGA-II} algorithms. It can be observed that all the optimal values converge to same value in all the three approaches.}
    \label{ab_duty}
\end{figure*}

Both the individual optimization problems were solved using genetic algorithms with polynomial mutation and cross-over probability of $0.9$. The payoff matrix for the Nash bargaining game is constructed using the best solutions $X_{1best}$ and $X_{2best}$. We compute $Y_1$ as well as $Y_2$ using $X_{1best}$ and $X_{2best}$. To compute $Y_{1_{X_{1best}}}$, $Y_{1_{X_{2best}}}$, $Y_{2_{X_{1best}}}$ and $Y_{2_{X_{2best}}}$ we pass $X_{1best}$ and $X_{2best}$ as features to the decision tree models $Y_1$ and $Y_2$.

\begin{align} \label{mat}
P = \begin{bmatrix}
 Y_{1_{X_{1_{best}}}} & Y_{2_{X_{1_{best}}}}\\
Y_{1_{X_{2_{best}}}} & Y_{2_{X_{2_{best}}}}
\end{bmatrix}
\end{align}

The disagreement values for the players is modeled as: 
\begin{align} 
   Y_{i_{worst}} = \min_{j = 1, 2} Y_{i_{X_{j_{best}}}}, \forall i \in \{1, 2\}
\end{align}

With respect to the furnace optimization, the matrix becomes,
\begin{align}
P = \begin{bmatrix}
 75.2 & 337.17\\
47.27 & 361.29
\end{bmatrix}
\end{align}

The diagonal values of the payoff matrix $P$ are the best solutions and off-diagonal values are the worst solutions. Worst value is the minimum guaranteed value for any player and the players start bargaining to get better values. Let $X_{best}$ be the best compromise solution which is acceptable for both the players and $Y_{1_{X_{best}}}$ and $Y_{2_{X_{best}}}$ be the pay-off values of the players (Absorbed Duty and COT). It follows that the optimization problem,(~\ref{super}) gives the best compromising solution,~\cite{10.5555/2655534}.

\begin{equation}
   \begin{aligned} \label{super}
   J = \text{max } (Y_{1_{X_{best}}}-Y_{1_{worst}})(Y_{2_{X_{best}}}-Y_{2_{worst}}) 
   \\
\text{subject to constraints:} \\
  Y_{1_{X_{best}}} \geq Y_{1_{worst}} \\
  Y_{2_{X_{best}}} \geq Y_{2_{worst}} \\  
\end{aligned} 
\end{equation}

From the matrix, it can be seen the worst values are:
\begin{align}
  Y_{1_{worst}} = 47.27\\
  Y_{2_{worst}} = 337.17
\end{align}

The players, as per the optimization problem (\ref{super}), try to maximize their values in such a way that the values are better than the worst values. This optimization problem was also solved using Genetic Algorithm \cite{ga} with same set of parameters as that of \textsc{NSGA-II} discussed in section \ref{nsga2}. The results of optimization are given by equations (\ref{Compromise}), (\ref{optimal1}) and (\ref{optimal2}):
\begin{align} 
\label{Compromise}
 X_{best} &= (83.41, 95.11, 202.85), \\ 
\label{optimal1}
Y_{1_{X_{best}}} &= 75.20, \\
\label{optimal2}
Y_{2_{X_{best}}} &= 361.29     
\end{align}

It can be observed that the best compromise solution values obtained after Nash bargaining (\ref{Compromise}) coincide with the optimum objective of the multi-objective optimization via both~\textsc{NSGA-II} as well as~\textsc{RNSGA-II} (\ref{nsga}).

\section{Comparison of Results} \label{results}
\begin{figure*}[h!]
    \centering
    \includegraphics[scale=0.275]{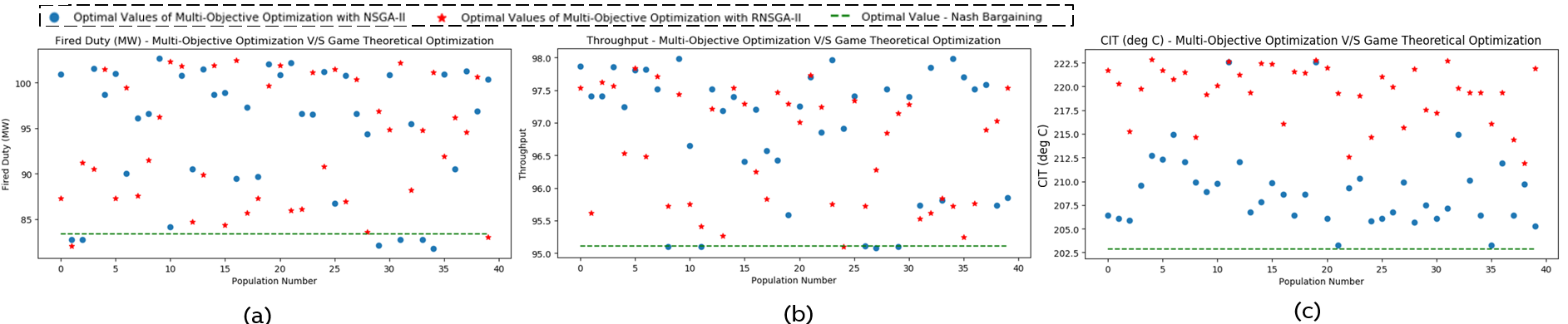}
    \caption{Comparison of optimal values of the manipulated variables obtained via Nash Bargaining, \textsc{NSGA-II} and \textsc{RNSGA-II} algorithms. The green dotted line shows the Nash Bargaining solution whereas the blue dots and the red stars indicate the optimal manipulated variable values of each population via \textsc{NSGA-II} and \textsc{RNSGA-II} algorithms. Both \textsc{NSGA-II} and \textsc{RNSGA-II} give multiple optimal configurations of the controlled variables. The Nash bargaining solution, on the contrary, is a single configuration of these variables.}
    \label{fd}
\end{figure*}
In this section, we analyze the results obtained by the two approaches and verify the effectiveness of the game theoretical approach. First, we compare the optimum objective values of Absorbed Duty and Coil Outlet Temperature from Figure (\ref{ab_duty}). Figures \ref{ab_duty}(a) and \ref{ab_duty}(b) compares the optimal values of Absorbed Duty and Coil Outlet Temperature. It is evident that both the algorithms \textsc{NSGA-II} and \textsc{RNSGA-II} give $40$ different candidate solutions all converging to a single value, which is same as the value obtained via Nash Bargaining. This is true for both the objectives.

Both \textsc{NSGA-II} and \textsc{RNSGA-II} give multiple optimal configurations of the manipulated variables Fired Duty, Throughput and CIT as seen from the figures \ref{fd}(a), \ref{fd}(b) and \ref{fd}(c). The Nash bargaining solution, on the contrary, is a single configuration of these variables. The optimal values of controlled variables (Absorbed Duty and COT) are same for all these values. From the figures, it is clear that the Nash Bargaining solution is comparable with the pareto-optimal solutions obtained via \textsc{NSGA-II} and \textsc{RNSGA-II} for all the three manipulated variables. The advantage of adopting the game methodology for process optimization is that we get a single best compromise solution, which aids the process engineer to set the manipulated variable values to this single value. He is not left with a further dilemma in deciding what solution to choose from the multiple optimal configurations generated by either \textsc{NSGA-II} or \textsc{RNSGA-II} algorithms, or any other multi-objective optimization algorithm.
\section{Conclusion and Future Work}


ML based models were used to study the effect manipulated variables on the controlled variables and the model outputs were used to perform optimization of furnace. Here, we used regression models to predict the controlled variable. However, the game theory based approaches can be used to integrate other advanced state-of-the art ML models, simulation models etc. The multi-objective nature of the furnace optimization problems makes them difficult to solve. Conventional evolutionary algorithms like \textsc{NSGA-II} and \textsc{RNSGA-II} were used to solve the multi-objective furnace optimization problem, however these algorithms resulted in Pareto-optimal frontiers which further pose questions on what decisions to make and what values of manipulated variables to be set from the multiple efficient configurations. As a solution to this, a cooperative game theory based approach via Nash Bargaining was implemented. The results of the game theory approach were then compared with that of the original solution of the problem solved using \textsc{NSGA-II} and \textsc{RNSGA-II} algorithms. From the results, it can be concluded that the cooperative game strategy gives results comparable with that of \textsc{NSGA-II} and \textsc{RNSGA-II}. This approach can be further extended to solve multi-objective optimization problems in the industries to find the best compromise solution among conflicting objectives. It eliminates the complications of multi-objective optimization and helps in faster decision making. 

We have considered limited set of manipulated and controlled variables for our experiments, due to data access limitations. In ideal setting, there are also other major factors such as air-fuel ratio, NOx and CO which are keys to control stack oxygen levels. Due to paucity of the downstream process details, we considered COT as a proxy target variable. However, typically it is decided by the product conversion in the downstream unit. In the present settings, throughput is considered as a manipulated variable as it can affect the upstream process. In present setup, we demonstrate the effectiveness of the cooperative games to optimize furnace efficiency under constrained settings. The complete formulation would require  better access to upstream and downstream processes, which in turn could change manipulated and controlled variables. The problem considered here is a minuscule component of the bigger plant optimization problem and is a step towards the Factory of Future. Complicated models to capture controlled variables might be essential, especially when optimization has to be done at plant level. Plant level optimization comprises of optimization at multiple levels. We are working on integrating deep learning models to solve multi-objective optimization problems, where we have more data with respect to a process.

\bibliographystyle{IEEEtran}
\bibliography{Main}

\begin{thebibliography}{10}
\providecommand{\url}[1]{#1}
\csname url@samestyle\endcsname
\providecommand{\newblock}{\relax}
\providecommand{\bibinfo}[2]{#2}
\providecommand{\BIBentrySTDinterwordspacing}{\spaceskip=0pt\relax}
\providecommand{\BIBentryALTinterwordstretchfactor}{4}
\providecommand{\BIBentryALTinterwordspacing}{\spaceskip=\fontdimen2\font plus
\BIBentryALTinterwordstretchfactor\fontdimen3\font minus
  \fontdimen4\font\relax}
\providecommand{\BIBforeignlanguage}[2]{{%
\expandafter\ifx\csname l@#1\endcsname\relax
\typeout{** WARNING: IEEEtran.bst: No hyphenation pattern has been}%
\typeout{** loaded for the language `#1'. Using the pattern for}%
\typeout{** the default language instead.}%
\else
\language=\csname l@#1\endcsname
\fi
#2}}
\providecommand{\BIBdecl}{\relax}
\BIBdecl

\bibitem{furnace}
M.~Masoumi and Z.~Izakmehri, ``Improving of refinery furnaces efficiency using
  mathematical modeling,'' \emph{International Journal of Modeling and
  Optimization}, 2011.

\bibitem{osti_1529693}
S.~Garcia and C.~T. Trinh, ``Comparison of multi-objective evolutionary
  algorithms to solve the modular cell design problem for novel biocatalysis,''
  \emph{Processes}, 2019.

\bibitem{MUSSATI20092194}
S.~Mussati, J.~I. Manassaldi, S.~J. Benz, and N.~J. Scenna, ``Mixed integer
  nonlinear programming model for the optimal design of fired heaters,''
  \emph{Applied Thermal Engineering}, 2009.

\bibitem{doi:10.1080/03052150902890064}
Y.~Hu and S.~S. Rao, ``Game-theory approach for multi-objective optimal design
  of stationary flat-plate solar collectors,'' \emph{Engineering Optimization},
  2009.

\bibitem{RAO1987119}
S.~Rao, ``Game theory approach for multiobjective structural optimization,''
  \emph{Computers and Structures}, 1987.

\bibitem{8680809}
B.~{Lokeshgupta} and S.~{Sivasubramani}, ``Cooperative game theory approach for
  multi-objective home energy management with renewable energy integration,''
  \emph{IET Smart Grid}, 2019.

\bibitem{article}
S.~Zara, A.~Dinar, and F.~Patrone, ``Cooperative game theory and its
  application to natural, environmental, and water resource issues: 2.
  application to natural and environmental resources,'' \emph{The World Bank,
  Policy Research Working Paper Series}, 2006.

\bibitem{5554307}
{Rui Meng}, {Ye Ye}, and {Neng-gang Xie}, ``Multi-objective optimization design
  methods based on game theory,'' in \emph{World Congress on Intelligent
  Control and Automation}, 2010.

\bibitem{pub.1101630721}
N.-g. Xie, Z.~Chen, K.~H. Cheong, R.~Meng, and W.~Bao, ``A multiobjective game
  approach with a preferred target based on a leader-follower decision
  pattern,'' \emph{Mathematical Problems in Engineering}, 2018.

\bibitem{pub.1042336139}
R.~Meng, N.~Xie, and L.~Wang, ``Multiobjective game method based on
  self-adaptive space division of design variables and its application to
  vehicle suspension,'' \emph{Mathematical Problems in Engineering}, 2014.

\bibitem{996017}
K.~Deb, A.~Pratap, S.~Agarwal, and T.~Meyarivan, ``A fast and elitist
  multiobjective genetic algorithm: Nsga-ii,'' \emph{Transactions on
  Evolutionary Computation}, 2002.

\bibitem{10.1145/1389095.1389231}
J.~Pfeiffer, U.~Golle, and F.~Rothlauf, ``Reference point based multi-objective
  evolutionary algorithms for group decisions,'' in \emph{GECCO}.\hskip 1em
  plus 0.5em minus 0.4em\relax Association for Computing Machinery, 2008.

\bibitem{pymoo}
J.~{Blank} and K.~{Deb}, ``Pymoo: Multi-objective optimization in python,''
  \emph{IEEE Access}, 2020.

\bibitem{10.5555/2655534}
Y.~Narahari, \emph{Game Theory and Mechanism Design}.\hskip 1em plus 0.5em
  minus 0.4em\relax World Scientific Publishing Company Pte. Limited, 2014.

\bibitem{ga}
P.~Bajpai and M.~Kumar, ``Genetic algorithm - an approach to solve global
  optimization problems,'' \emph{Indian Journal of Computer Science and
  Engineering}, 2010.

\end{thebibliography}

\end{document}